\documentclass[aps,prb,superscriptaddress,twocolumn,tightenlines]{revtex4-1}
\usepackage{amssymb}
\usepackage{graphicx}
\usepackage{natbib}
\usepackage{epstopdf}
\usepackage{dcolumn}% Align table columns on decimal point
\usepackage{bm}% bold math
\usepackage{makeidx}
\usepackage{enumerate}
\usepackage {graphicx}
\usepackage {amsmath}
\usepackage {amsfonts}
\usepackage {amssymb}
\usepackage {mathrsfs}
\usepackage {bbm}
\usepackage{subfigure}
\begin{document}
\title{Time-reversal symmetry breaking state in dirty three-band superconductor}

\author{Valentin Stanev}
\affiliation{Condensed Matter Theory Center, Department of Physics,
University of Maryland, College Park, MD 20742-4111, USA}

\date{\today }

\begin{abstract}
I study the effects of disorder on the superconductivity of a three-band model with repulsive interband pairing. Such a  model can support several possible superconducting order parameters, including a complex time-reversal symmetry breaking (TRSB) state. Impurity scattering suppresses the critical temperature of all these states, but the complex state survives, and remains a part of the phase diagram of the model even in the presence of moderate amount of disorder. This means that the TRSB states could be experimentally accessible in multiband materials like iron pnictides and chalcogenides.

\end{abstract}
\maketitle

\section{Introduction}
The discoveries of MgB$_2$\cite{MgB} and the iron-based superconductors\cite{LaOFeAs} have demonstrated  that multiband superconductivity is a phenomenon with far reaching  fundamental and applied implications. It provides a new route to high-temperature superconductivity, and for engineering superconducting materials with novel properties. Research in this area has been very active, and, as a result, rich physics has been uncovered. One of the most exciting ideas that have been discussed is the possibility of time-reversal symmetry breaking (TRSB) states in multiband systems \cite{sid1, sid2, sid3, sid4,Agterberg, VS, Tanaka2, Hu,Dias, Orlova, Maiti,Babaev2, Benfatto, Lin}. They appear as complex admixture of  distinct and competing superconducting order parameters, like $s$-wave and $d$-wave. The three-band system with repulsive interband pairing is the prototypical model which exhibits intrinsic frustration: the presence of several superconducting states with the same symmetry, which are directly competing with each other. In this model there are three possible two-gap order parameters with sign change between the gaps of each possible pair ($s_{\pm}$ states). They cannot be easily reconciled with each other, and the complex TRSB state appears as a natural compromise for a three-gap order parameter. This model is thought to be relevant for the iron-based superconductors, most of which are believed to be in some form of the $s_{\pm}$ state (see recent reviews\cite{Review,Reviews1,Reviews2}). The TRSB state can potentially be tuned by doping in some of these materials \cite{Maiti}.

Most of the research on TRSB states was done in the clean limit, ignoring the effects of the (unavoidable) impurities and defects. However, it has been long recognized that even nonmagnetic impurities can
be detrimental to multiband superconductivity \cite{SungJPCS67,Muzikar,
GM1, Kulic, Mishonov, Gurevich}, especially in the case of an order parameter with
sign change (for example, $d$-wave or $s_{\pm}$ state)\cite{Muzikar, Senga, Bang,
Vorontsov}. Effects of disorder on the TRSB states have not been studied so far, and it remains unclear if these states are reasonably robust, or, on the contrary, fragile and likely to be absent in any real material. In this paper I address this question, and consider the simplest possible three-band model with repulsive pairing interaction, which has well-established complex state in the clean limit. Introducing disorder, I study its effects on the critical temperature line and the possible phases close to it. I demonstrate that $T_c$ is strongly suppressed by disorder. The complex state, however, is still possible, almost until superconductivity is completely destroyed. This surprising, at first sight, robustness of the TRSB state is, in fact, easy to understand. As mentioned, this state is a complex admixture of three order parameter -- the real two-gap $s_{\pm}$ states. Even though suppressed, these real states survive  moderate amount of disorder. As long as they are present, there is always a region on the phase diagram where they compete, and in it the complex state emerges as a compromise that minimizes the free energy. Thus, the TRSB state, arising in the tree-band model, could be observable in real materials (in which some degree of disorder is aways present). In contrast, for cases in which one of the competing phases is more susceptible to destruction by disorder than the others (say $d$-wave vs. $s$-wave), it seems likely that the TRSB state region can be completely wiped out.   
%Unlike the $s$-wave superconductivity in a single-band materials, protected by the Anderson theorem from non-magnetic impurities.
%They can dramatically change the properties of the system, such as density of
%states, $T_c$ and  $H_{c2}$.

It is interesting to note that disorder-induced interband scattering can
continuously change the order parameter of a two-band superconductor from $s_{\pm}$
to $s_{\mathrm{++}}$\cite{Golubov, Golubov2, GM2} state. 
This transition can involve an intermediate complex combination of these two order parameters\cite{VS2}, and also leads to TRSB state. It is effect entirely due to the presence of interband impurity scattering. Thus, in some cases disorder is a necessary ingredient for the TRSB states. Furthermore, similar role to that of disorder can be played by a boundary scattering\cite{Bobkovi}, or a proximity-coupled gap\cite{Ng,VS3,VS4}

\section{Three-band system with interband pairing}

I consider a system described by the following Hamiltonian:
\begin{eqnarray}
\mathscr{H} &= & \sum_{i,\bf{k},\sigma} \xi^{(i)}_{\bf{k}} c^{(i)\dagger}_{\bf{k}\sigma} c^{(i)}_{\bf{k}\sigma}  + \sum_{i,j,\bf{k, k'},\sigma} u^{imp}_{ij} c^{(i)\dagger}_{\bf{k}\sigma} c^{(j)}_{\bf{k'}\sigma} + \nonumber \\ 
&+& \sum_{i, j,\bf{k, k'}} u^{sc}_{ij} c^{(i)\dagger}_{\bf{k}\uparrow} c^{(i)\dagger}_{\bf{-k}\downarrow} c^{(j)}_{\bf{k'}\uparrow} c^{(j)}_{\bf{-k'}\downarrow} 
+ h.c. .
\label{Hamiltonian}
\end{eqnarray}
Here $i$ and $j$ are band indicies, running from $1$ to $3$.
The bands are parabolic, with partial and total densities of states (DOS) $N_1$, $N_2$, $N_3$ and $N_t=N_1+N_2+N_3$ respectively. For simplicity I assume they are identical, i.e., $N_1=N_2=N_3=N_0$. 

The first term in Eq. \ref{Hamiltonian} is the kinetic energy of the electrons, while the second term is due to impurities, which introduce  both intraband and interband scattering. The scattering rates are parametrized by $\gamma_{ij}$. The
interband terms ($i \neq j$) can be written as $\gamma=N_0 \Gamma$, with
$\Gamma=n_{imp}\pi (u^{imp})^2$, where $n_{imp}$ and $u^{imp}$ are the impurities' concentration
and  potential respectively (assuming $u_{imp}$ is the same for each $(i,j)$ combination).
%AEKOct13:
Since point defects, such as atomic substitutions or vacancies, can
scatter carriers with large momentum change, the intraband and interband scattering rates are expected to be comparable \cite{Kemper}. 

Superconductivity is driven by the pairing potentials $u^{sc}_{ij}$, which can scatter Cooper pairs within the same band ($i=j$), or between different bands ($i\neq j$). Note that the Cooper pairs themselves are always composed of electrons from the same band. In general, there are  six independent pairing interactions. For simplicity, I ignore the inraband terms $u^{sc}_{ii}$. Furthermore, I reduce the remaining three interband terms $u^{sc}_{ij}$ to two, by assuming that $u_{12}=u_{13}$. Finally, the interaction potentials are converted to dimensionless constants by using the DOS:
\begin{eqnarray}
\lambda_{12} &=& N_0 u^{sc}_{12},\  \lambda_{13} = N_0 u^{sc}_{13},\  \lambda_{23} = N_0 u^{sc}_{23}; \nonumber\\
\lambda_{12}&=&\lambda_{13} \equiv \lambda,\ \  \lambda_{23}\equiv \eta;\ \ \lambda, \eta > 0 ~,\nonumber
\end{eqnarray}    
The last inequality is true for systems with only repulsive interband pairing interactions. All the coupling constants can be combined into a $3\times 3$ symmetric matrix 
\begin{eqnarray}
\hat{\lambda}\equiv \left(
\begin{matrix}
  0     &\lambda  &\lambda  \\ 
\lambda &     0   &\eta    \\ 
\lambda & \eta    &0 
\end{matrix}
\right).
\label{lambda}
\end{eqnarray}
I study the phase diagram of the model, assuming that $\eta$ is fixed, and varying $\lambda$ from $0$ (only two bands are coupled) to $\lambda\gg\eta$ (a pair of bands is much weaker coupled). Of special interest is the region around the degeneracy point $\lambda=\eta$. 

In the superconducting state there are three gap parameters
$\Delta_1$, $\Delta_2$ and $\Delta_3$, describing the superconducting condensate in each band. They can be written as
$\Delta_i = |\Delta_i | e^{i\varphi_i}$; the phases $\varphi_i$ are gauge-dependent, but their differences $\phi_{ij}=\varphi_i- \varphi_j$ are physically meaningful,
gauge-invariant quantities. Note that the superconducting state breaks only the overall $U(1)$ symmetry. 

The system with $\hat{\lambda}$ given in Eq. \ref{lambda} has obvious frustration: each pair of gaps $(i,j)$ prefers to have  $\phi_{ij}=\pi$, due to the repulsive $\lambda_{ij}$ pairing. However, since there are three such pairs, there is no way to satisfy this preference for all gaps simultaneously. Thus, the different possible solutions compete, and this competition is the source of the TRSB state. 

In the clean limit the gap parameters obey three coupled nonlinear self-consistency equations:
\begin{equation}
\Delta_i=- \lambda_{ij} \pi T \sum_{\omega>0}^{\omega_0} \frac{\Delta_j}{\sqrt{\omega^2 + |\Delta_j|^2}},
\label{gap_eq1}
\end{equation}  
where $\omega=\pi T(2n + 1)$ is the Matsubara frequency, and $\omega_0$ is a high-energy cut-off (e.g., the Debye frequency). When disorder is present, it leads to additional interband scattering and coupling of the gaps, and the self-consistency equations become considerably more complicated (see, e.g., Ref. \onlinecite{Maki}). However, close to the critical temperature these equations can be simplified by expanding them in powers of
$|\Delta_i|$. In the presence of impurities  this can be done systematically, starting from the Usadel equations\citep{Gurevich, AEK}. 

\begin{figure}[h]
\begin{center}$
\begin{array}{cc}
\includegraphics[width=0.22\textwidth]{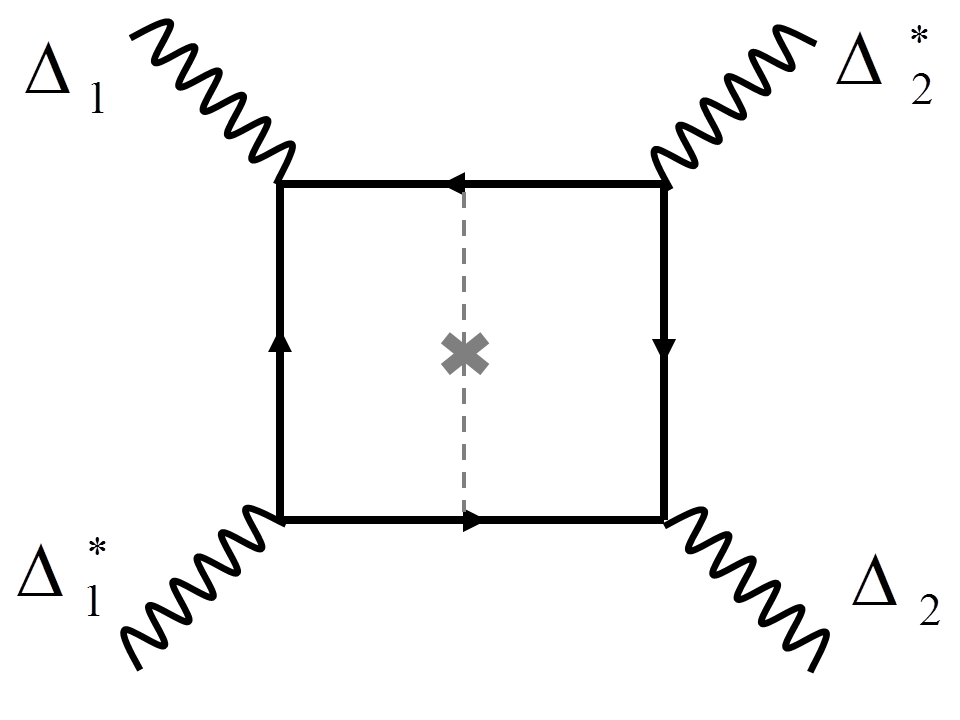}
\includegraphics[width=0.22\textwidth]{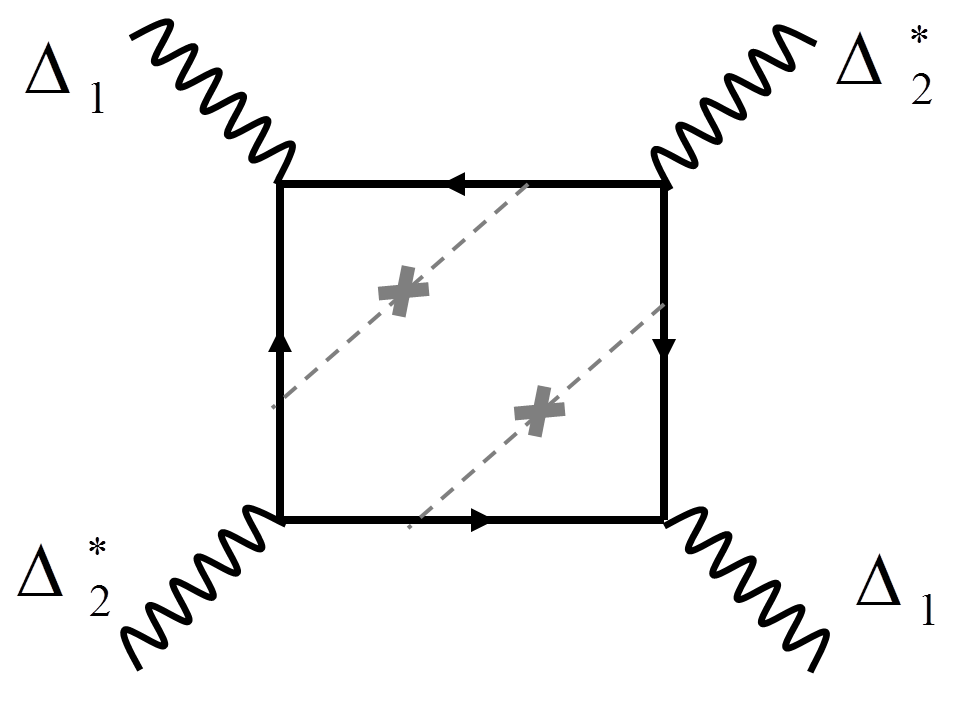}
\end{array}$
\end{center}
\caption{Diagrams of processes contributing to the coefficients $b_{12}$ (on the left), and $c_{12}$ (on the right) of $\mathcal{F}_{ij}$. The solid lines are electrons, and the dashed line represent the interband impurity scattering.}
\label{Fig1}
\end{figure}

\begin{figure}[h]
\begin{center}$
\begin{array}{cc}
\includegraphics[width=0.22\textwidth]{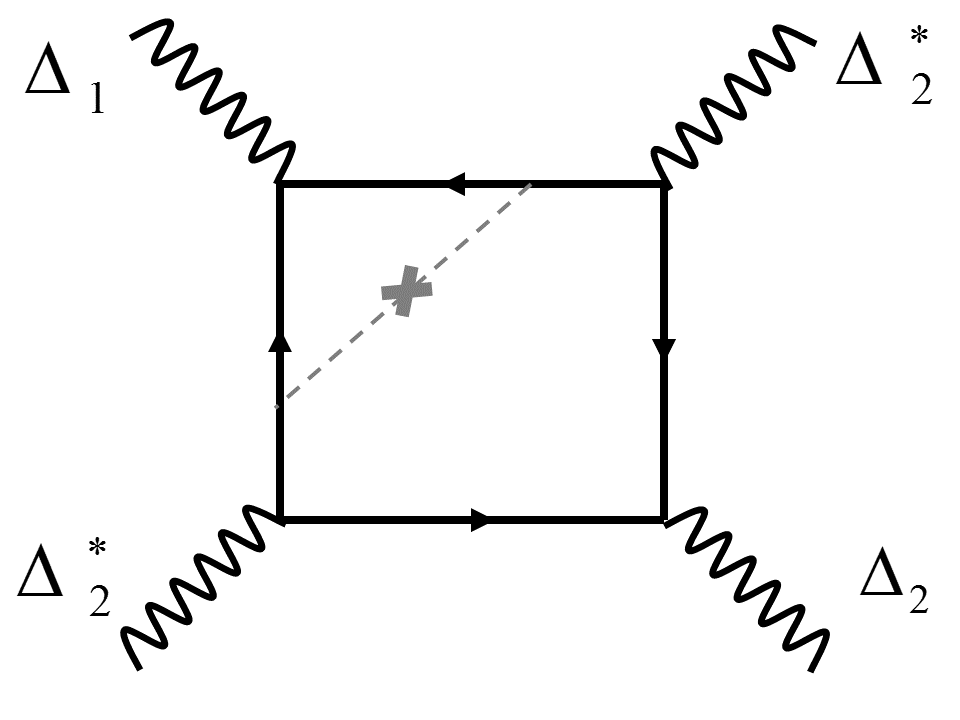}
\includegraphics[width=0.22\textwidth]{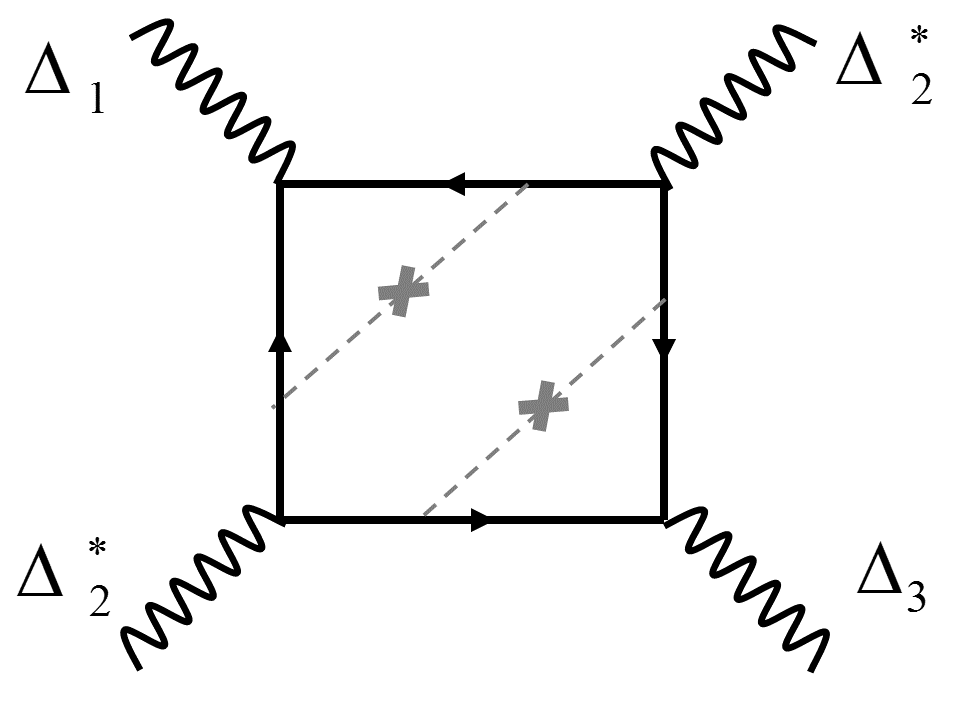}
\end{array}$
\end{center}
\caption{Diagrams of processes contributing to the coefficients $c_{22}$ (on the left), and $d_{213}$ (on the right). The meaning of the lines is the same as in Fig. \ref{Fig1}.}
\label{Fig1a}
\end{figure}

The resulting gap equations can be though of as derived from an effective multigap Ginzburg-Landau (GL) free energy. In the clean case this free energy looks relatively simple \cite{Tilley, ZD}. 
%(In the clean case this has been done starting from the gap equations \cite{Tilley, ZD}. However, the precise meaning and validity of the multiband extension is a matter of ongoing debate\cite{AEK, Babaev1, Kogan1, Shanenko,GLnote}.)
 When interband impurity scattering is added, theory becomes more complicated; for the two-gap case it was derived earlier \cite{VS2}. With three gaps the free energy can be written as
\begin{eqnarray}
\mathcal{F}_{GL}= \mathcal{F}_{i} + \mathcal{F}_{ij}+ \mathcal{F}_{ijk}+\mathcal{F}_{EM},
\label{GL1}
\end{eqnarray}
where we have kept only terms up to quartic order in $\Delta$. The intraband terms $\mathcal{F}_{i}$ look like three copies of standard single-gap GL theory,
\begin{eqnarray}
 \mathcal{F}_{i}(\Delta_i) = a_{i}|\Delta_{i}|^2 + \frac{b_{i}}{2}|\Delta_{i}|^4, % + K_{m}|\Pi\Delta_{m}|^2.
\label{GL2}
\end{eqnarray}
but with coefficients modified by disorder. In contrast, in $\mathcal{F}_{ij}$ there are many new terms appearing due to scattering from impurities. It can be written as:
\begin{eqnarray}
 \mathcal{F}_{ij}(\Delta_i, \Delta_j) = a_{ij}|\Delta_{i}||\Delta_{j}|\cos{\phi_{ij}} + b_{ij}|\Delta_{i}|^2|\Delta_{j}|^2\nonumber\\ +
 c_{ii}|\Delta_{i}|^3|\Delta_{j}|\cos{\phi_{ij}} +c_{ij}|\Delta_{i}|^2|\Delta_{j}|^2\cos{(2\phi_{ij})}.
\label{GL3}
\end{eqnarray}
 For processes that contribute to these terms see Fig. \ref{Fig1} and Fig. \ref{Fig1a}\cite{GLnote2}. In the clean limit ($\gamma_{ij}\rightarrow 0$) the standard GL function for multiband superconductor is recovered: $\mathcal{F}_{ij} \rightarrow a_{ij}|\Delta_{i}||\Delta_{j}|\cos{\phi_{ij}}$. Notice that $\mathcal{F}_{ij}$ is a simple extension of the two-band theory\cite{VS2}.
In contrast, the $\mathcal{F}_{ijk}$ term contains entirely new, intrinsically three-gap contributions:
\begin{eqnarray}
 \mathcal{F}_{ijk}(\Delta_i, \Delta_j, \Delta_k) = d_{ijk}|\Delta_{i}|^2|\Delta_{j}||\Delta_{k}|\cos{(\phi_{ij}+ \phi_{ik})}.
\label{GL4}
\end{eqnarray}
This term is solely due to impurities, and it disappears in the clean limit (see Fig. \ref{Fig1a}). 

$\mathcal{F}_{EM}$ combines the electromagnetic field energy, and the
derivative terms that couple $\Delta_i$ to the electromagnetic
vector-potential. For the rest of this paper  uniform order parameter is assumed, and no field, so I will ignore $\mathcal{F}_{EM}$.

It is important to emphasize that the microscopically derived mean-field equations   are the main object of our theory, and the free energy  given in Eq. \ref{GL1} is secondary. Although it may seem unnecessary to distinguish them,  there are at least two problems in taking $\mathcal{F}_{GL}$ as a starting point. First, the region of validity of the multiband GL theory is a matter of ongoing debate\cite{AEK, Babaev1, Kogan1, Shanenko,GLnote}. In addition, and even more seriously, when the interband pairing dominates, this simple GL theory becomes unstable. This is due to the presence of a passive superconducting channel, which has to be treated properly\cite{Benfatto}. The mean-field equations, however, remain valid.

\section{At the $T_c$ line}
At $T_c$ only the linear terms in the self-consistency equations are important, and the problem can be reduced to a simple matrix one:
\begin{eqnarray}
\mathbb{M} \hat{\Delta}=-\hat{\Delta},
\end{eqnarray}
where I have defined the column vector $\hat{\Delta}$ (with transpose $\hat{\Delta}^T=(\Delta_1, \Delta_2, \Delta_3)$). $\mathbb{M}$ is a $3 \times 3$ symmetric matrix:
\begin{eqnarray}
\mathbb{M}\equiv
\begin{bmatrix}
       2 \lambda I_{\gamma} & \  \lambda I_2\! + 2\! \lambda I_{\gamma} & \lambda I_2\! + 2\! \lambda I_{\gamma} 
       \\[0.3em]
       \lambda I_2\! + 2\! \lambda I_{\gamma} & \  (\lambda+\eta) I_{\gamma} & \eta I_2 + (\lambda+\eta) I_{\gamma}
      \\[0.3em]
      \lambda I_2\! + 2\! \lambda I_{\gamma}  &  \eta I_2 +(\lambda+\eta) I_{\gamma}& (\lambda+\eta) I_{\gamma}
            \\[0.3em]
     \end{bmatrix}\nonumber.
\end{eqnarray}
Here I have introduced  
\begin{eqnarray}
I_1= 2 \pi T \sum_{0}^{\omega_0}\frac{1}{|\omega_n|},\ \ I_2= 2 \pi T \sum_{0}^{\omega_0}\frac{1}{|\omega_n|+ 2 \gamma}, \nonumber
\end{eqnarray}
and $I_{\gamma}=I_1 - I_2$. 
%Here $\omega_0$ is a high-energy cut-off (e.g., the Debye frequency).
The matrix is written in a form that separates the effects of the interband impurity scattering. In the clean limit we have $I_{2} \rightarrow I_1=\gamma_0 \ln(2 \omega_0/\pi T)$, (with $\gamma_0$ being the Euler constant), and $I_{\gamma} \rightarrow 0$, thus recovering the equations obtained earlier\cite{VS}. Also, note that the intraband scattering rate does not appear at all -- this is a consequence of the Anderson theorem\cite{Anderson}.
 \begin{figure}[h]
 \begin{center}$
 \begin{array}{cc}
 \includegraphics[width=0.43\textwidth]{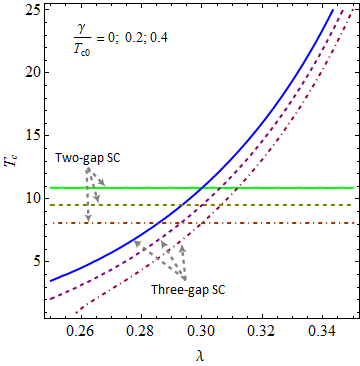}
 \end{array}$
 \end{center}
 \caption{$T_c$ for the two-gap and the three-gap order parameters as a function of the coupling constant $\lambda/\eta$ for three  disorder strengths $\gamma/T_{c0}$: $0$ (solid line); $ 0.2$ (dashed line); $ 0.4$ (dashed-dotted line). The critical temperature of the two-gap state does not depend on $\lambda$. $T_{c0}$ is the value of $T_c$ for the two-gap state without disorder.}
 \label{Fig2}
 \end{figure}
The eigenvectors of $\mathbb{M}$ represent the possible  order parameters that can condense, and the corresponding eigenvalues determine their respective transition temperatures. There are two possible order parameters, as in the clean case, with the third eigenvector having unphysical transition temperature (as already mentioned, the presence of this passive superconducting channel has important consequences\cite{Benfatto}). Quite surprisingly, the form of $\hat{\Delta}$ at $T_c$ does not depend on $\gamma$ \emph{at all}, and we recover the clean limit order parameters exactly:
\begin{eqnarray}
\hat{\Delta}^{(1)}\propto
\begin{bmatrix}
      0 
       \\[0.3em]
       -1
      \\[0.3em]
      1 
            \\[0.3em]
     \end{bmatrix},\ \ \
 \hat{\Delta}^{(2)}\propto
 \begin{bmatrix}
       -\frac{\eta + \sqrt{\eta^2+ 8 \lambda^2}}{2 \lambda}
        \\[0.3em]
        1
       \\[0.3em]
       1 
             \\[0.3em]
      \end{bmatrix}.    
\end{eqnarray}
 The first eigenvector represents a two-gap solution, in which there is a sign change between the two bands that become superconducting; the third band stays normal. The other eigenvector has gaps on all three bands, with two of them having the same sign. These solutions represent different compromises for the underlying frustrated system.
      
The critical temperatures do depend, of course, on $\gamma$, and are given by:
\begin{eqnarray}
\frac{1}{\eta}=I_{2}, \ 1=(\eta + 2 \lambda -\sqrt{\eta^2 + 8 \lambda})I_{\gamma} + \frac{\eta - \sqrt{\eta^2+ 8 \lambda^2}}{2} I_2. \nonumber
\end{eqnarray}
%In the clean limit we have $I_2=\gamma_0 \ln(2 \omega_0/\pi T)$, (with $\gamma_0$ being the Euler constant). 

\begin{figure}[h]
\begin{center}$
\begin{array}{cc}
\includegraphics[width=0.43\textwidth]{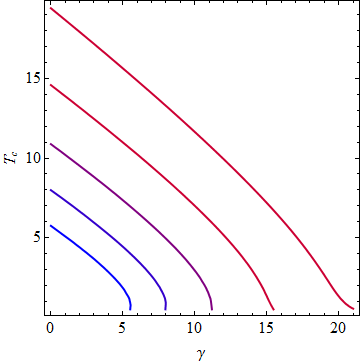}
\end{array}$
\end{center}
\caption{$T_c$ for the three-gap order parameter as a function of the disorder strength for five values of $\lambda/\eta$ (in order of increasing $T_{c0}$): $0.90$ (blue);  $0.95$ (blue-purple); $1$ (purple); $1.05$ (purple-red); $1.10$ (red).  The $T_c$ lines for the two-gap and three-gap states coincide for $\lambda=\eta$ (purple line). $T_{c0}$ for the two-gap case is the same as on Fig. \ref{Fig2}.}
\label{Fig3}
\end{figure}

So how do the impurities affect the critical temperature? We plot the $T_c$ lines for both superconducting states as function of $\lambda$ for several values of $\gamma$ (shown on Fig. \ref{Fig2}). As we can see, the two states' critical temperatures cross at $\lambda=\eta$ for any disorder strength. For  $\lambda$ smaller than $\eta$ the two-gap solution is the leading instability (since the system prefers to keep the weaker-coupled band normal). For $\lambda>\eta$ the three-gap state condenses first (and the weaker-coupled pair of bands has relative phase $\phi_{23}=0$). So far, this is completely analogous to the clean case. The only effect of impurity scattering is to suppress $T_c$ of both states, and eventually a critical amount of disorder  destroys the superconductivity in the system. This is as expected: because we are considering only repulsive interactions, superconductivity depends crucially on the sign change in the order parameter. Since the interband scattering tends to average the gaps on the different bands, large enough $\gamma$ averages the order parameter down to zero. Plot of $T_c$ vs. $\gamma$ is shown on Fig. \ref{Fig3}. While for the two-gap state the $T_c$ suppression follows the universal Abrikosov-Gor'kov (AG) curve\cite{AG}, the three-gap case appears more complicated, with clear deviation from the AG curve at very low critical temperatures and large $\lambda/\eta$ ratios. However, when $\lambda$ is close to $\eta$ the disorder suppression of the two critical temperatures is (almost) equal.   

At the point $\lambda=\eta$ the two-gap and the three gap solutions are degenerate, and any linear combination of them is a possible order parameter. In particular, a complex state can be constructed \cite{Agterberg,VS}:  $(\hat{\Delta}^{(3)})^T= \Delta_0 (1, e^{2 \pi i/3}, e^{-2 \pi i/3})$. It was demonstrated in the clean limit that below $T_c$ this state plays an important role in the region of maximal frustration of the system $\lambda \approx \eta$. As we will see in the next section, this remains true in the presence of disorder.

\section{TRSB state}
It was shown in the preceding section that there are two possible order parameters that can condense. Which one is leading depends on the ratio $\lambda/\eta$.  Below $T_c$ the situation becomes considerably more complicated, and when $\lambda \approx \eta$ the system may prefer a complex admixture of the two real states. To study this possibility we write a general complex order parameter
$(\hat{\Delta}^{(3)})^T=\Delta_0 (- \kappa, e^{i \varphi}, e^{-i \varphi}) $ (it incorporates the $2\leftrightarrow 3$ symmetry of the problem). Note that $\varphi= \phi_{23}/2$. By appropriate choice of gauge $\Delta_1$ is made real (which is always possible). Unlike the other two order parameters, however, $\hat{\Delta}^{(3)}$ is intrinsically complex (for $\kappa\neq 0$). This signifies the spontaneous breaking of the time-reversal symmetry of the system by the order parameter. It is caused by chiral pair currents in momentum space, when a pair is scattered from one band to another in a specific pattern (for example $1\rightarrow 2\rightarrow 3$). This symmetry breaking requires a second phase transition, below the customary superconducting one (associated with the $U(1)$ symmetry).

It is important to understand the difference between the $s_{++}+ i s_{\pm}$ states in the two-band\cite{VS2} and the three-band systems\cite{VS}. In the former the complex state is driven by interband impurity scattering, which leads to a quartic $\cos(2 \phi_{ij})$ term in the GL expansion (a term that vanishes in the clean limit)\cite{VS2}. In contrast, the three-band case has an intrinsic instability towards a TRSB state, due to the competition between the three quadratic $\cos(\phi_{ij})$ terms (see, e.g., Ref. [\onlinecite{Maiti}]). This explains the different behavior of the two systems close to $T_c$ (compare the phase diagrams constructed in Refs. [\onlinecite{VS}] and [\onlinecite{VS2}]).

The intrinsic instability of the three-band model can be used to simplify the calculation. Close to $T_c$ the most important terms in the self-consistency equations are linear (i.e., the terms with coefficients $a_{i}$ and $a_{ij}$). Thus, we can concentrate on the effects of impurities on those terms, and neglect all impurity induced terms, which are small anyway. (This is equivalent to only keeping $a_{ij}$ from $\mathcal{F}_{ij}$, and completely ignoring $\mathcal{F}_{ijk}$.\cite{GLnote3})  

%We derive the GL equations by minimizing $\mathcal{F}_{GL}$ with respect to $\varphi$ , $\kappa$, and $\Delta_0$.
% Retaining only $a_{ij}$ from $\mathcal{F}_{ij}$, and neglecting $\mathcal{F}_{ijk}$ altogether we obtain from  $\partial \mathcal{F}_{GL}/\partial \varphi=0$:
With this, the first self-consistency equation is: 
 \begin{equation}
 \cos{\varphi}=\frac{\eta \kappa}{2 \lambda}\left(\frac{1+ 2 \lambda I_{\gamma}}{1+ 2 \eta I_{\gamma}}\right).
 \label{Eq_varphi} 
 \end{equation}
% The equation $\partial \mathcal{F}_{GL}/\partial \kappa=0$ is 
The other two equations are more illuminating if written in terms of the GL coefficients $a_i$ and $a_{ij}$, where we have $a_1=\eta/(2 \lambda^2)- I_{2}$, $a_2=a_3=1/(2 \lambda)- I_{2}$,  $a_{12}=a_{13}=-(1/\lambda +2 I_{\gamma})$, and $a_{23}=-(1/\eta + 2 I_{\gamma})$. (Note that $a_{ij}$ are negative, which seems to directly contradict the idea of phase frustration. It is, however, just a sign that the na\"{i}ve GL theory is unstable, and more care is needed in order to construct a proper GL free energy in the case interband pairing dominates\cite{Benfatto}.) The remaining mean-field equations are  
 \begin{equation}
-\kappa\left(a_{1}- \frac{a_{12}^2}{2 a_{23}}\right)= 2 b_{1} \Delta_0^2 \kappa^3,
\label{Eq_kappa}
 \end{equation}
and
 \begin{equation}
- a_1(\kappa^2+2)+ \kappa^2\frac{a_{12}^2}{2 a_{23}} + \cos(2 \varphi) a_{23} = 2   b_{1}(\kappa^4 +2) \Delta_0^2.
\label{Eq_Delta}
  \end{equation}
Alternatively, these equations can be obtained from $\partial \mathcal{F}_{GL}/\partial \varphi=0$, $\partial \mathcal{F}_{GL}/\partial \kappa=0$ and  $\partial \mathcal{F}_{GL}/\partial \Delta_0=0$.

Analyzing these equations 
%Eqs. \ref{Eq_varphi} and \ref{Eq_kappa}
provides a lot of useful information. First, note that they reduce correctly to the clean limit results \cite{VS} for $I_{\gamma}\rightarrow 0$. From Eq. \ref{Eq_varphi} we see that nonzero $\kappa$ leads to $\varphi \neq \pi/2$. Thus, if we start  from two-gap state ($\kappa=0$, $\varphi = \pi/2$) the system can continuously transform \emph{only} to the TRSB state, but not  directly to the three-gap one. Also, note from Eq. \ref{Eq_kappa} that the two-gap solution with $\kappa=0$ is always possible. However, if the right side is positive the equation can have also non-trivial $\kappa$ solution. Since $a_{23}<0$ the second term in the brackets is negative. On the other hand, $a_{1}$ is positive at high temperatures, and only changes sign and turns negative for finite temperature below $T_c$. Thus, the expression in the brackets can turn negative at some $T$ below the critical temperature, and the TRSB state with $\kappa\neq 0$ and $\varphi \neq \pi/2$ emerges.
\begin{figure}[h]
\begin{center}$
\begin{array}{cc}
\includegraphics[width=0.43\textwidth]{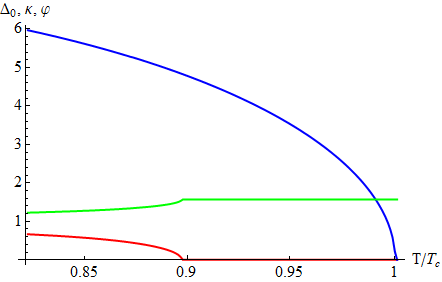}
\end{array}$
\end{center}
\caption{$|\Delta_0(T)|$ (blue), $\kappa(T)$ (red), and $\varphi(T)$ (green) for $\lambda/\eta=0.99$, with $\gamma/T_{c0}=0.3$ (moderate disorder). Immediately below $T_c$ the order parameter is in the two-gap state ($\kappa=0$, $\varphi=\pi/2$), but transitions to TRSB at $T\approx 0.9 T_c$.}
\label{Fig4}
\end{figure}

\begin{figure}[h]
\begin{center}$
\begin{array}{cc}
\includegraphics[width=0.43\textwidth]{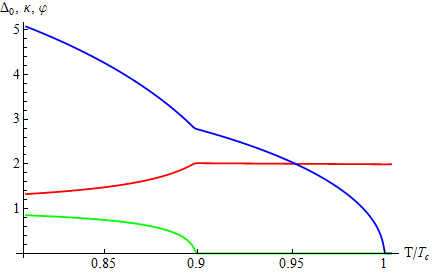}
\end{array}$
\end{center}
\caption{The same as on Fig.\ref{Fig4}, but for $\lambda=1.01\eta$. The real three-gap state condenses first, but there is second transition to TRSB state around $0.9 T_c$.}
\label{Fig5}
\end{figure}

To demonstrate this, I solved Eqs. \ref{Eq_varphi}, \ref{Eq_kappa} and \ref{Eq_Delta} numerically, on both sides of the degeneracy point $\lambda=\eta$. Results are shown on Figs. \ref{Fig4} and \ref{Fig5}. For $\lambda<\eta$ the two-gap state is the first to condense, and for $\lambda>\eta$ the three-gap state is leading. The TRSB state appears below $T_c$ in both cases, despite the presence of disorder in the system. Note that the disorder is relatively weak, and the $T_c$ suppression is small --  only about $30$ percent in both cases. However, calculations show the  that the TRSB state survives even in the case of much stronger disorder, when the critical temperature is only a small fraction of the clean limit $T_c$, provided that both real order parameters still have non-zero $T_c$. Eventually, disorder completely suppresses the weaker superconducting channel, at which point TRSB state also disappears. However, as long as the two channels are nearly degenerate (i.e., $\lambda \approx \eta$), this single-order-parameter region is small, and TRSB state survives as a part of the phase diagram.

\section{Discussion and Conclusion}

In this paper I considered the influence of disorder on a simple three-band model with repulsive interband pairing interactions. This model is known to support a complex TRSB superconducting state in the region where the competition between the real two-gap and three-gap order parameters is strongest. As expected, the presence of impurities tends to suppress the critical temperature of all states, and for some finite value of disorder superconductivity is completely destroyed. However, the complex state remains possible in parts of the phase diagram even in the presence of moderate disorder, as long as there are two real superconducting channels. 

The considered model and the derived results have to be taken with some caution. The three-band system is quite simple, and does not represent well the situation in iron-based superconductors (which, at the very least, have more bands at the Fermi level). This simplicity is the reason behind some artificial results, like the presence of ungapped band in the two-gap phase. This phase would likely disappear in a more realistic, and less symmetric cases, in agreement with the general argument in Ref. \onlinecite{Barzykin}. Furthermore, the results are derived using expansion of the mean-field equations, valid only close to the critical temperature. In addition, I have neglected some quartic terms, with the assumption that they will not change the structure of the phase diagram (at least close to $T_c$).  In spite of the simplifications, these results are clear indication that TRSB states, if present in the iron-based superconductors, can be relatively robust against disorder, and thus experimentally accessible. 

The discussion in this paper has been mostly concerned with the stability of the TRSB state. Once it is present, equally important and interesting questions are of the physical properties and the possible manifestations of this state. Its presence requires a superconductor-to-superconductor transition, which may be detected in bulk thermodynamic property like the specific heat. Once inside the complex state, its time-reversal nature can lead to induced currents and spontaneous magnetization close to impurities and edges \cite{sid3} (in the bulk the breaking is a consequence of interband Josephson currents, which do not lead to observable magnetization). Collective modes of the TRSB state are also quite interesting, as they consist of mixed phase-amplitude fluctuations of {\it different} gaps (see, for example, Refs. \onlinecite{Carlstrom,Lin2,VS5, Benfatto}). Another interesting consequence of the discreet symmetry breaking is the possibility of domain walls, whose presence can be used as a signature of this state \cite{Garaud}.

\section{Acknowledgments}
I acknowledge many useful discussions with Alexei Koshelev. This research was supported by DOE-BES DESC0001911, and Simons Foundation.

 \bibliographystyle{apsrev}

\begin{thebibliography}{10}

\bibitem{MgB} J. Nagamatsu, N. Nakagawa, T. Muranaka, Y. Zenitani, and J. Akimitsu, Nature \textbf{410}, 63 (2001).

\bibitem{LaOFeAs}Y. Kamihara, T. Watanabe, M. Hirano, and H. Hosono, J. Am.
Chem. Soc. \textbf{130}, 3296 (2008).

%\bibitem{physicaC}P. C. W. Chu \textit{et al.} (eds.), \textit{\ Superconductivity in iron-pnictides}, Physica C \textbf{469} (special issue), 313-674 (2009).


\bibitem{sid1} Y. Ren, J.-H. Xu, and C. S. Ting, Phys. Rev. B \textbf{53}, 2249 (1996).

\bibitem{sid2} K. A. Musaelian, J. Betouras, A. V. Chubukov, and R. Joynt, Phys. Rev. B \textbf{53}, 3598 (1996).

\bibitem{sid3} W.-C. Lee, S.-C. Zhang, and C. Wu, Phys. Rev. Lett. \textbf{102}, 217002 (2009).

\bibitem{sid4} C. Platt, R. Thomale, C. Honerkamp, S.-C. Zhang, and W. Hanke,   Phys. Rev. B \textbf{85}, 180502 (2012).


\bibitem{Agterberg} D. F. Agterberg, V. Barzykin, and L. P. Gor'kov, Phys. Rev. B \textbf{60}, 14868 (1999).

\bibitem{VS} V. Stanev and Z. Te\v sanovi\' c, Phys. Rev. B \textbf{81},
134522 (2010).

\bibitem{Tanaka2} Y. Tanaka and T. Yanagisawa, Sol. State. Commun. \textbf{150}, 1980 (2010).

\bibitem{Hu} X. Hu and Z. Wang, Phys. Rev. B \textbf{85}, 064516 (2012).

\bibitem{Dias} R. Dias and A. Marques, Supercond. Sci. Technol. \textbf{24}, 085009 (2011).

\bibitem{Orlova} N. V. Orlova, A. A. Shanenko, M. V. Milosevic, F. M. Peeters, A. Vagov, and V. M. Axt,  Phys. Rev. B \textbf{87}, 134510 (2013).

\bibitem{Maiti} S. Maiti and A. V. Chubukov, Phys. Rev. B \textbf{87}, 144511 (2013).

\bibitem{Babaev2} T. Bojesen, E. Babaev, and A. Sudbo,Phys. Rev. B \textbf{88}, 220511(R) (2013).

\bibitem{Benfatto} M. Marciani, L. Fanfarillo, C. Castellani, and L. Benfatto, Phys. Rev. B \textbf{88}, 214508 (2013).

\bibitem{Lin} For a review, see, for example, Shi-Zeng Lin, arXiv:1408.5938 (2014).

\bibitem{Review}J-P. Paglione and R. L. Green, Nature Phys. \textbf{6}, 645 (2010).

\bibitem{Reviews1}P. J. Hirschfeld, M. M. Korshunov, and I. I. Mazin, Rep. Prog. Phys. \textbf{74}, 124508 (2011);

\bibitem{Reviews2} A. V. Chubukov, Annu. Rev. Cond. Mat. Phys. \textbf{3}, 57 (2012).

\bibitem{SungJPCS67} C.\ C.\ Sung and V.\ K.\ Wong, J.\ Phys.\ Chem.\ Sol.,
\textbf{\ 28}, 1933 (1967).

\bibitem{Muzikar} G. Preosti, H. Kim, and P. Muzikar, Phys. Rev. B \textbf{50}, 13638 (1994).

\bibitem{GM1} A. A. Golubov and I. I. Mazin, Phys. Rev. B \textbf{55}, 15146 (1997).

\bibitem{Kulic} M.L. Kulic and O.V. Dolgov, Phys. Rev. B \textbf{60}, 13062 (1999).

\bibitem{Mishonov} T. Mishonov, E. Penev, J. Indekeu, and V. Pokrovsky, Phys. Rev. B \textbf{68}, 104517 (2003).

\bibitem{Gurevich} A. Gurevich, Phys. Rev. B \textbf{67}, 184515 (2003).


\bibitem{Senga} Y. Senga and H. Kontani, J. Phys. Soc. Jpn. \textbf{77}, 113710 (2008).

\bibitem{Bang} Y. Bang, H.-Y. Choi, and H. Won, Phys. Rev. B \textbf{79}, 054529 (2009).

\bibitem{Vorontsov} A.B. Vorontsov, M.G. Vavilov, and A.V. Chubukov, Phys. Rev. B \textbf{79}, 140507(R) (2009).

\bibitem{GM2} A. A. Golubov and I. I. Mazin, Physica C \textbf{243}, 153 (1995).

\bibitem{Golubov} D. V. Efremov, M. M. Korshunov, O. V. Dolgov, A. A. Golubov, and P. J. Hirschfeld,    Phys. Rev. B \textbf{84}, 180512(R) (2011).

\bibitem{Golubov2} D.V. Efremov, A.A. Golubov, O.V. Dolgov,  New J. Phys. \textbf{15}, 013002 (2013).

\bibitem{VS2} V. Stanev and A. E. Koshelev,  	Phys. Rev. B \textbf{89}, 100505(R) (2014).

\bibitem{Bobkovi} A. M. Bobkov and I. V. Bobkova, Phys. Rev. B \textbf{84}, 134527 (2011).

\bibitem{Ng} T. K. Ng and N. Nagaosa, Euro. Phys. Lett. \textbf{87}, 17003 (2009).

\bibitem{VS3} A. E. Koshelev and V. Stanev, Europhys. Lett. \textbf{96}, 27014 (2011).

\bibitem{VS4}V. Stanev and  A. E. Koshelev, Phys. Rev. B \textbf{86}, 174515 (2012).

\bibitem{Kemper} See, for example, A. F. Kemper, C. Cao, P. J. Hirschfeld, and H.-P. Cheng, Phys. Rev. B \textbf{80}, 104511 (2009).

\bibitem{Maki} K. Maki, Superconductivity, ed. R. D. Parks, Vol. 2 (Dekker, New York, 1969).

\bibitem{AEK} A. E. Koshelev and A. A. Golubov, Phys. Rev. Lett. \textbf{92}, 107008 (2004).

\bibitem{Tilley} D.R. Tilley, Proc. Phys. Soc. \textbf{84}, 573 (1964).

\bibitem{ZD} M. E. Zhitomirsky and V.H. Dao, Phys. Rev. B \textbf{69}, 054508 (2004).

\bibitem{GLnote2} Explicit expressions for $a_{ij}$, $b_{ij}$, and $c_{ij}$ for arbitrary $\gamma$ and $T$ are provided in the Supplemental Material of Ref. \onlinecite{VS3}. Closed forms for these coefficients in terms of the digamma function and its derivatives can also be obtained. 

\bibitem{Babaev1} E. Babaev and J. M. Speight, Phys. Rev. B \textbf{72}, 180502 (2005).

\bibitem{Kogan1} V. G. Kogan and J. Schmalian, Phys. Rev. B \textbf{83}, 054515 (2011).

\bibitem{Shanenko} A. A. Shanenko, M. V. Milosevic, F. M. Peeters, and A. V. Vagov, Phys. Rev. Lett. \textbf{106}, 047005 (2011).

\bibitem{GLnote} Since I am mostly concerned with the case when two superconducting channels are nearly degenerate, the singe-order-parameter description\cite{Kogan1} shrinks to a very narrow region close to $T_c$, and one has to use the full three-gap equations.

\bibitem{Anderson} P. W. Anderson, J. Phys. Chem. Solids \textbf{11}, 26 (1959).

\bibitem{AG} A. A. Abrikosov and L. P. Gor'kov, Zh. Eksp. Teor. Fiz. \textbf{39}, 1781 (1960) [Sov. Phys. JETP \textbf{12}, 1243 (1961)].

\bibitem{GLnote3} Numerical calculations show that, in the physically relevant regime, the quartic, intrinsically two-band, contributions (e.g. $c_{ij}$) are at least an order of magnitude smaller then the single band ones ($b_{ii}$). This justifies dropping them in the usual GL region of validity (close to $T_c$). In it these terms are sub-leading, in the sense that the only result of their presence would be to move the boundaries of the different phases slightly, and would not change the overall topology of the phase diagram.

\bibitem{Barzykin} V. Barzykin, Phys. Rev. B \textbf{79}, 134517 (2009).

\bibitem{Carlstrom} J. Carlstrom, J. Garaud, and E. Babaev, Phys. Rev. B, \text{84}, 134518 (2011).

\bibitem{Lin2} S.-Z. Lin, and X. Hu, Phys. Rev. Lett. \text{108}, 177005 (2012).

\bibitem{VS5} V. Stanev, Phys. Rev. B \text{85}, 174520 (2012).

\bibitem{Garaud} J. Garaud, and E. Babaev, Phys. Rev. Lett. \textbf{112}, 017003 (2014).

     

\end{thebibliography}

\end{document}